\documentclass{llncs}
\usepackage{graphicx}
 \usepackage[ruled,vlined]{algorithm2e}
 \usepackage{cite}

\title{ A Fast Fault Tolerant Partitioning Algorithm for Wireless Sensor Networks}
\author{Dibakar Saha\inst{1} \and Nabanita Das\inst{2}}
\institute{Advanced Computing and Microelectronics Unit\\Indian Statistical Institute\\ 203 B.T. Road, Kolkata-700108, India}

\begin{document}
\maketitle \begin{center}
\email{{\scriptsize \inst{1}dibakar.saha10@gmail.com , \inst{2}ndas@isical.ac.in }}
\end{center}

\begin{abstract}

In this paper, given a random uniform distribution of sensor nodes on a 2-D plane, a fast self-organized distributed algorithm is proposed to find the maximum number of partitions of the nodes such that each partition is connected and covers the area to be monitored.
Each connected partition remains active in a round robin fashion to cover the query region individually.
In case of a node failure, the proposed distributed fault recovery algorithm reconstructs the affected partition locally utilizing the available free nodes.
Simulation studies show significant improvement in performance compared to the earlier works in terms of computation time, the diameter of each partition, message overhead and network lifetime.

\keywords{Wireless Sensor Network (WSN), Self Organization, Coverage, Partition, Network Lifetime, Fault Tolerant}

\end{abstract}

\vspace{-0.73cm}
\section{Introduction}
\vspace{-0.2cm}
\label{intro}
In an over deployed Wireless Sensor Network (WSN), a large number of sensor nodes are randomly deployed to monitor a large geographical area. Each sensor node is integrated with processing elements, memory, battery power and wireless communication capabilities. Once deployed, they are, in general left unattended. Hence due to power drainage, hardware degradation, environmental hazards etc. sensor nodes are much prone to failures. For better utilization of the over-deployed nodes to save energy and to extend the lifetime of the network, this paper addresses the problem of finding maximum number of partitions of the sensor nodes such that each partition is connected and covers the whole query region. Instead of keeping all the sensors active always, these partitions will remain active one after another in a round robin fashion. Therefore, if there are $K$ such partitions, the network lifetime will be enhanced by at most $K$ times. Here, given a random uniform distribution of sensor nodes over a 2-D plane, a
distributed algorithm is developed for finding the maximum number of partitions of connected nodes such that each partition ensures coverage. In case of node failures, a distributed algorithm is
developed for fault recovery that rearranges the affected partition locally to tolerate single node faults within a partition.
 Simulation studies show that compared to the earlier techniques, the proposed algorithm is faster and results better partition topology with reduced diameter and requires less message overhead. Also, in case of unpredictable node faults the neighboring nodes execute the localized fault recovery algorithm that rearranges the partition locally to make the system fault-tolerant. Simulation results show that it extends the network lifetime significantly.

The rest of the paper is organized as follows. Section \ref{sec:rel_work} presents a brief outline of related works. Section \ref{sec:proposed_model} describes the proposed model. Section \ref{sec:algo} includes the proposed algorithms. Simulation results are described in Section \ref{sec:results} and Section \ref{sec:conclusion} concludes the paper.
% \vspace{-0.5cm}
\vspace{-0.35cm}
\section{Related Works}
\vspace{-0.3cm}
\label{sec:rel_work}
Extensive research results have been reported so far addressing the problems of sensing coverage and network connectivity in wireless sensor networks. In many works, the authors considered only the coverage issues in wireless sensor networks.
In\cite{ Li, Meguerdichian}, authors propose efficient distributed algorithms to optimally solve the coverage problem in WSN.
In \cite{ Demin}, authors provide an analytical framework for the coverage problem and lifetime maximization of a WSN.
 In \cite{ jie} a decentralized and localized node density control algorithm is proposed for network coverage.
The work in\cite{Huang} proposed three approximation algorithms for the {\it set-k Cover} problem, where each point of the query region will be covered by at least $k$ nodes.
The work in \cite{ SSlij} considers the problem of maximizing the number of disjoint sets of sensor nodes to cover the query region. \\
But unless the coverage and connectedness problems are considered jointly, the data sensed by the nodes covering the region can not be gathered at the sink node in multi hop WSN's.
Authors of \cite{ Himangsu}, \cite{ Zhou} focused on both connectivity and coverage problems with the objective of finding a single connected set cover only.
The problem of finding a connected set cover of minimum size is itself an {\it NP-hard} problem \cite{ Himangsu}. Some of the papers considered the fault tolerant connected set cover problems.
 An approximation algorithm is proposed in \cite{Zhang} for fault tolerant connected set cover problem.
In \cite{ Peng}, a coverage optimization algorithm based on particle swarm optimization technique is developed.
In \cite{Lin,Chong,Tian,Wang,Gallais}, authors proposed several dynamic localized algorithms to achieve the coverage and connectivity. But dynamic algorithms, in general, require large message overhead to collect current neighborhood information at some intervals. Also, finding just a single connected cover keeps a large number of sensors unutilized. Hence, the authors in \cite{ Pervin}, propose a localized algorithm for finding maximum number of connected set covers that is to be executed once during network initialization only. In some papers \cite{ Gallais, Pervin, Tian}, it has been assumed that the query area is a dense grid \cite{ Wei}, \cite{ RSS} composed of unit cells. The knowledge of exact location of each node is needed here. A sensor node computes the covered area by counting the cells covered by each neighbor that makes the procedure computation intensive. To avoid this, in \cite{ dibakar} the {\it DCSP} algorithm is proposed where authors assume that the monitoring area is divided
into a limited number of square blocks such that a sensor node within a block completely covers it irrespective of its position within the block.Therefore, the coverage problem can be solved
easily with much less computation. However, the proposed distributed algorithm was a slow one requiring $p$ rounds to achieve a partition with $p$ nodes. Also, the fault model considers the faults due to energy exhaustion only that assumes that a node can predict its failure and can inform its neighbors in advance.\\
 In this paper, a faster distributed algorithm requiring less message overhead is proposed that is executed during network initialization only. It attempts to create maximum number of connected partitions of sensor nodes with reduced diameter such that each partition covers the area under investigation and being active in round robin fashion it enhances the network lifetime manifold. The reduced diameter of the partition keeps the communication latency low. Moreover, a distributed fault recovery algorithm is developed for a stronger fault model that in presence of any unpredictable node faults, can rearrange the affected partition locally, so that it remains operational. Simulation studies show that this fault recovery scheme enhances the network lifetime by more than $50\%$.
% \vspace{-0.5cm}
\vspace{-0.4cm}
\section{Proposed Model and Problem Formulation}
\begin{figure}[ht]
\begin{minipage}[b]{0.5\linewidth}
\centering
 \includegraphics[scale=0.4]{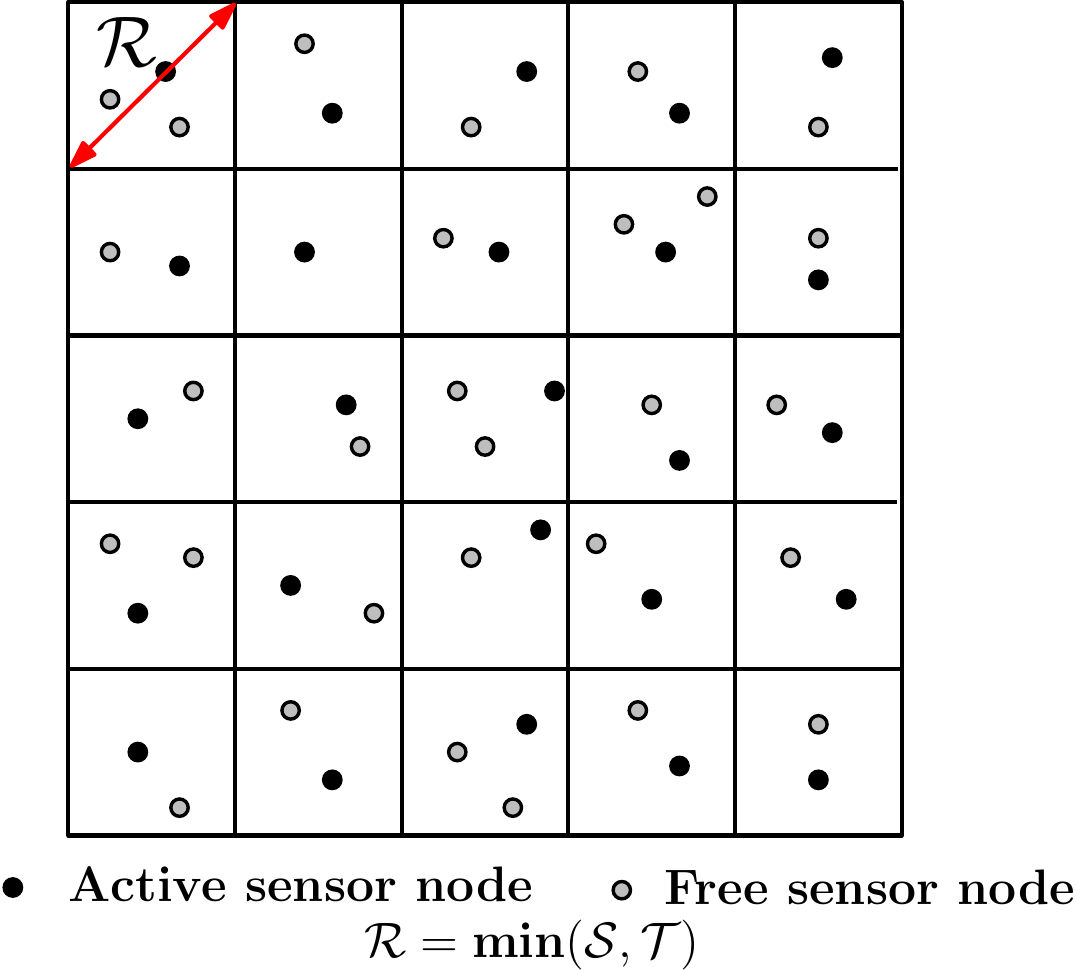}
\caption{Nodes in P divided into a grid of square blocks}
 \label{fig1}
 \vspace{-0.1cm}
 \end{minipage}
%  \hspace{0.1cm}
 \begin{minipage}[b]{0.5\linewidth}
\centering
 \includegraphics[scale=0.4]{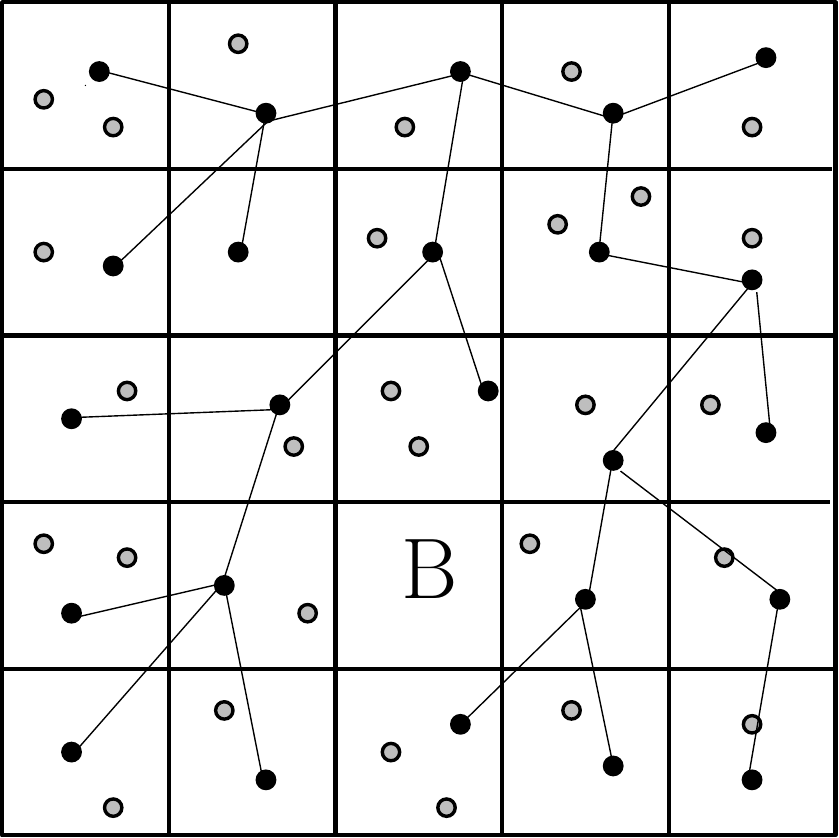}
 \vspace{0.8cm}
\caption{Connectivity without coverage}
 \label{fig:coverage}
\end{minipage}
\end{figure}

\vspace{-0.3cm}
 \label{sec:proposed_model}
Let $n$ homogeneous sensor nodes be deployed over a 2-D region $P$ each with same sensing range $\cal S$ and transmission range $\cal T$. It is assumed that $P$ is divided into a finite number of square blocks \cite{  dibakar}. Each side of the block is $\frac {\cal R}{\surd 2}$ as shown in Fig.\ref{fig1}, where $\cal {R}$= min$(\cal S,T)$.
\vspace{-0.5cm}
\vspace{-0.5cm}
Therefore, it is evident that each sensor node completely covers the block it belongs to and all nodes within the same block are connected to each other. Hence, activating just a single sensor node from each block is sufficient to cover the region $P$. But it is not guaranteed that any such set is connected or not.
\begin{figure}[ht]
 \begin{minipage}[b]{0.45\linewidth}
 \centering
 \includegraphics[scale=0.4]{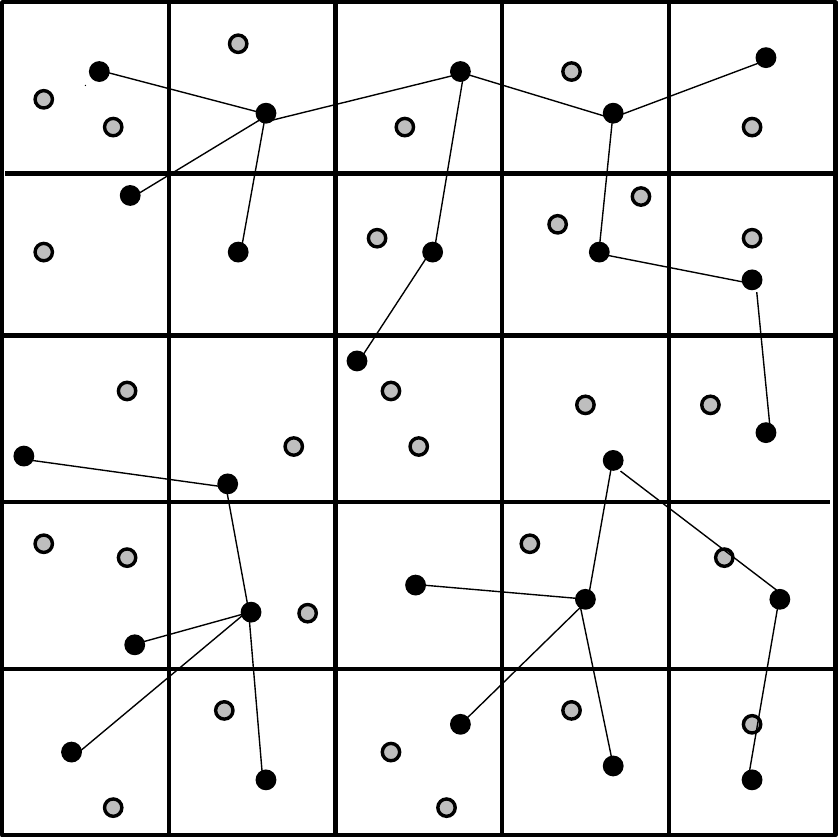}
\caption{Coverage without connectivity}
 \label{fig:connectivity}
 \vspace{-0.1cm}
 \end{minipage}
  \begin{minipage}[b]{0.5\linewidth}
\centering
 \includegraphics[scale=0.4]{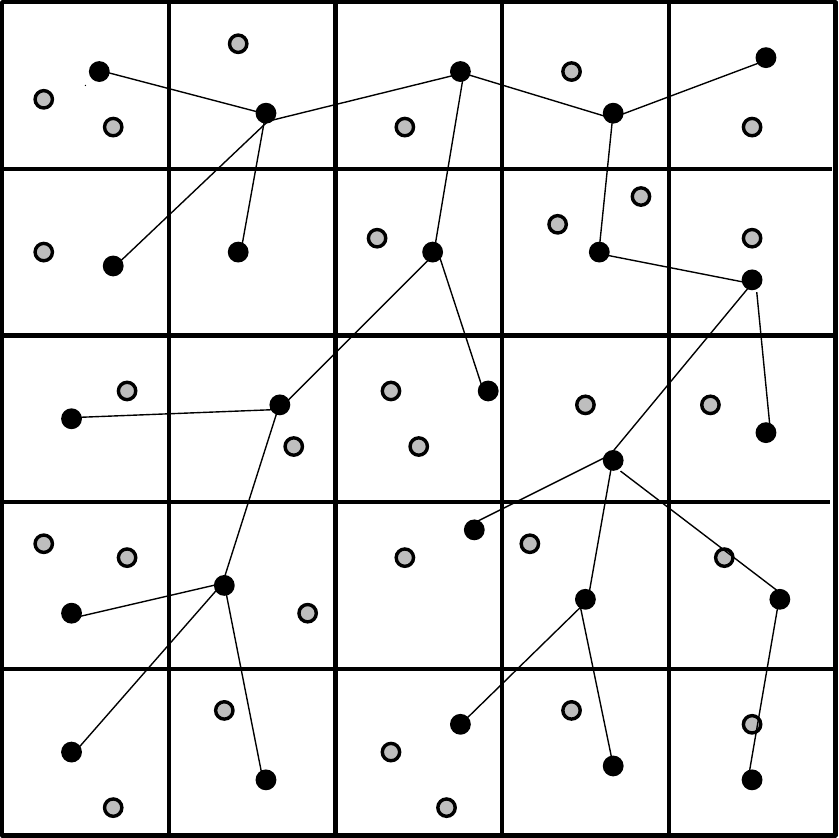}
\caption{Coverage with connectivity}
 \label{fig:coverage_connectivity}
 \vspace{-0.1cm}
 \end{minipage}
\end{figure}
As for example, Fig.\ref{fig:coverage} shows a partition where the selected nodes are connected but a block B is not covered. Whereas, Fig. \ref{fig:connectivity} shows a partition where all blocks are covered but the nodes are not connected, and finally, Fig. \ref{fig:coverage_connectivity} shows the desired topology where the partition covers all blocks as well as it is connected.
Assuming this grid structure of the query region $P$, this paper addresses the {\it connected set cover partitioning } problem introduced in \cite {Pervin}. For completeness, the problem is defined below.
\begin{definition}
 Consider a sensor network consisting of a set $S$ of $n$ sensors and a query region $P$. A set of sensors $N\subseteq S$ is said to be a {\emph connected $1$-cover} for $P$ if, each point $p\in P$ is covered by at least one sensor from $N$ and the communication graph induced by $N$ is connected.
 \vspace{-0.2cm}
\end{definition}
\paragraph{\bf Connected Set Cover Problem} Given $n$ sensor nodes distributed over a query region, the {\emph Connected Set Cover Problem} is to find a {\it connected $1$-cover} of minimum size. This problem is known to be an NP-hard problem \cite{Himangsu}.
% \vspace{-0.4cm}
\paragraph{\bf Connected Set Cover Partitioning Problem} The Connected Set Cover Partitioning Problem is to partition the sensors into {\it connected $1$-covers} such that the number of covers is maximized \cite{Pervin}.

The following section describes the algorithms developed for solving the Connected Set Cover Partitioning Problem.
% \vspace{-0.2cm}
\vspace{-0.3cm}
\section{Algorithm for Connected Cover Partitioning}
\vspace{-0.2cm}
\label{sec:algo}
In pervasive computing environments, it is evident that in most of the cases the system captures data in distributed nodes communicating through poorly connected network. Since, in WSN large number of sensor nodes are deployed over a geographical area, to collect information of the whole network at a central node is not feasible in terms of message overhead and energy requirement.
Instead, it is better to compute in a distributed fashion based on the local neighborhood information using less communication.
Hence the focus of our work is on distributed computation of the connected partitions.\\
In this section, a distributed algorithm is developed to find the maximum number of {\it connected-$1$ covers} of a WSN. Also, in the presence of a node fault, a localized algorithm is presented to rearrange the affected partition to make the system fault tolerant.
 \vspace{-0.3cm}
\subsection{Distributed Algorithm for Partitioning}
It is assumed that a set of $n$ sensor nodes $S=\{s_1, s_2, s_3, \ldots, s_n\}$ is deployed on a 2-D plane $P$ divided into say,
$m$ square blocks $P$= $\{ p_1, p_2, p_3, \ldots, p_m\}$, as has been described in Section 3.
Each square block has unique id. Each sensor node knows the location in terms of its block within which it is located.\\
We propose the following types of messages to be exchanged among nodes.
\begin{itemize}

 \item { \bf Selectlist($C_i, i, \{j\}$) :} This message is sent by a node-$i$ that selects a list of neighbors $\{j\}$ for inclusion in its partition with leader $C_i$.

 \item {\bf Selected($C_i, \{j\}$) :} This message is initiated by the leader node $C_i$ and is sent to node-$\{j\}$ for inclusion in its partition.

\item {\bf Confirm($C_i,j$) :} Node $j$ sends this message to the leader after joining the partition $C_i$.

\item {\bf Include($C_i,j$) :} Leader $l$ broadcasts this message within $C_i$ to include node-$j$ in $C_i$.
\end{itemize}
Depending on the node density, a probability value $0 < l_{prob} < 1$ is determined to select $p$ number of leader nodes randomly.
Each node $i\in S$ generates a random number $r$ to check if $r \leq l_{prob}$, the {\it leader probability}.
If yes, it becomes a leader node and sets its parent as null.
If $p$ leader nodes emerge, ${\cal L}=\{ l_1, l_2, l_3,\ldots, l_ p\}$, each leader initiates the creation of one partition concurrently to generate disjoint connected set covers.
 \vspace{-0.5cm}
 \begin{figure}[ht]
\begin{minipage}[b]{0.3\linewidth}
\centering
\includegraphics[scale=0.35]{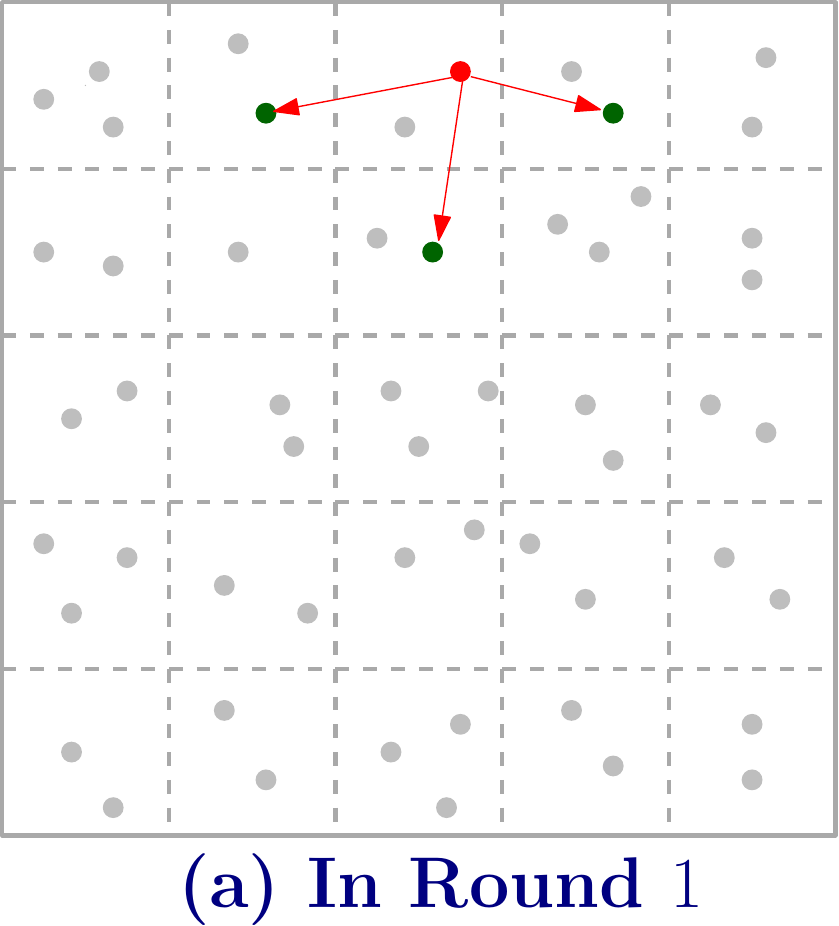}
% \hspace{0.63cm}
\end{minipage}
\begin{minipage}[b]{0.3\linewidth}
\centering
\includegraphics[scale=0.35]{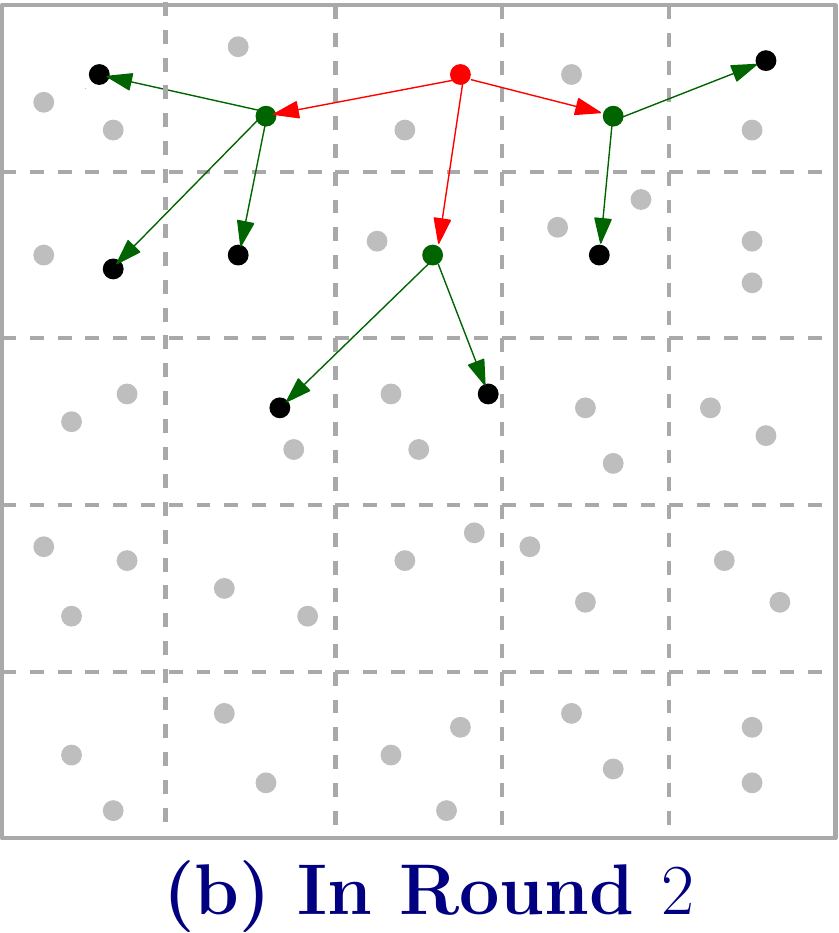}
 \end{minipage}
 \begin{minipage}[b]{0.3\linewidth}
\centering
\includegraphics[scale=0.35]{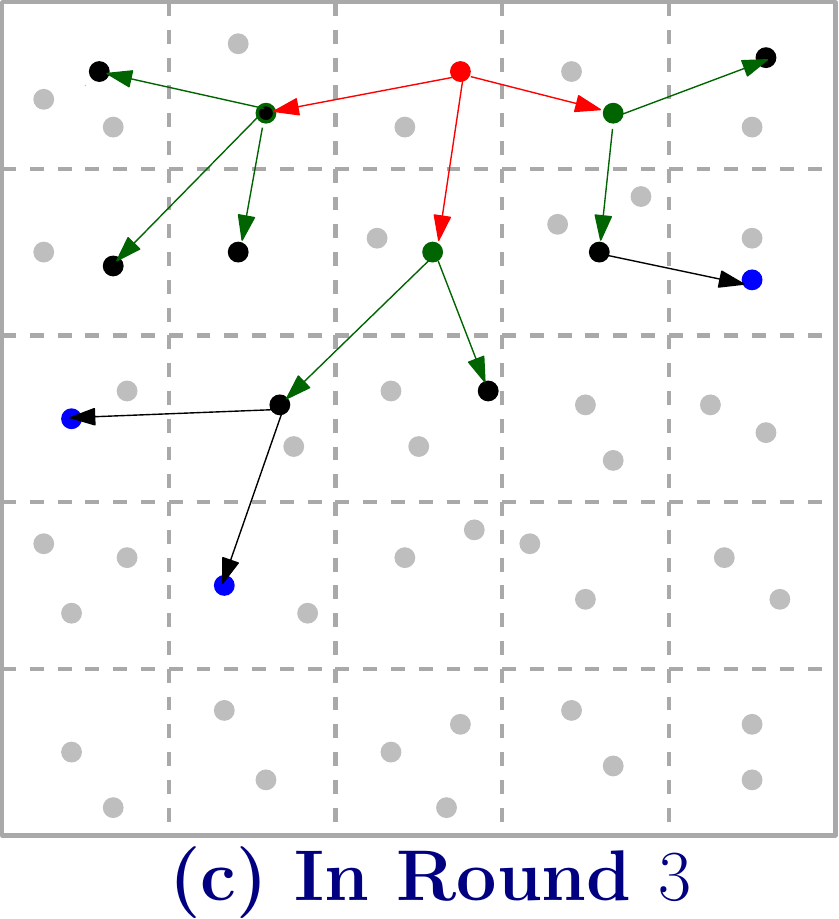}
 \end{minipage}
\hspace{0.5cm}
 \begin{minipage}[b]{0.3\linewidth}
\centering
\includegraphics[scale=0.35]{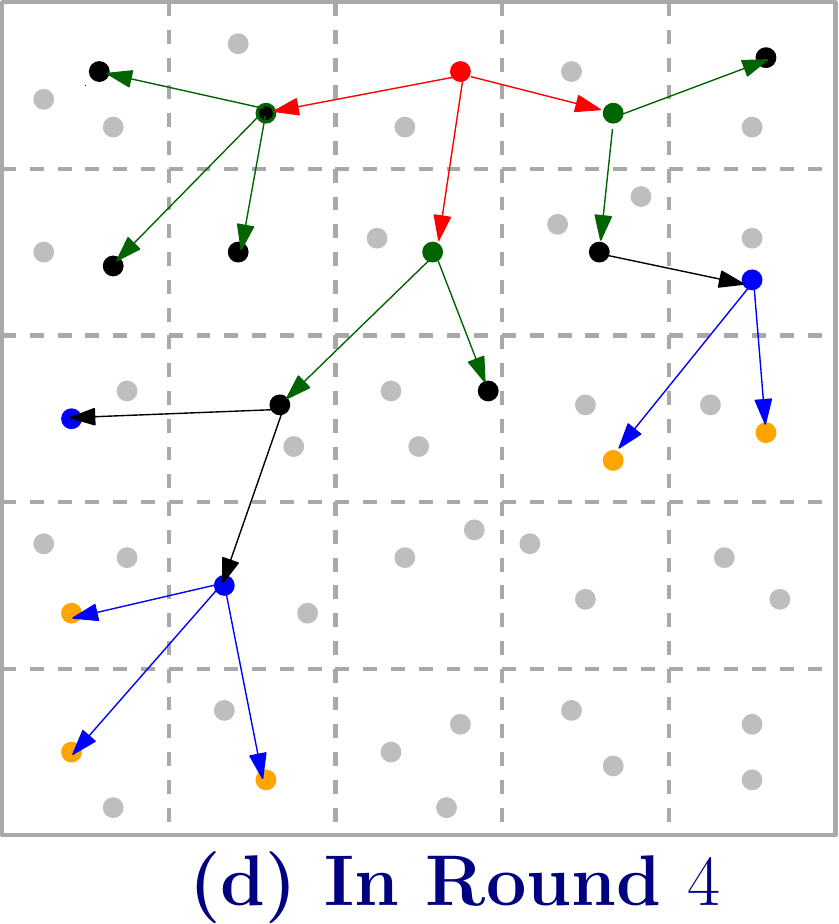}
 \end{minipage}
 \hspace{0.4cm}
\begin{minipage}[b]{0.3\linewidth}
\centering
\includegraphics[scale=0.35]{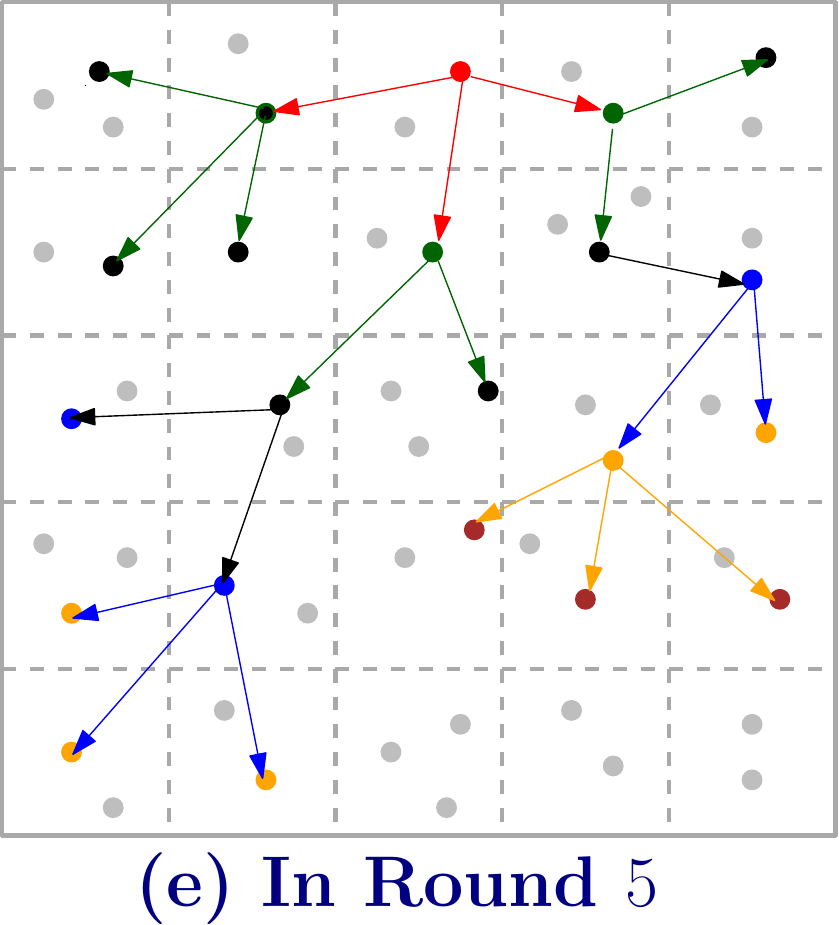}
 \end{minipage}
 \hspace{0.4cm}
 \begin{minipage}[b]{0.3\linewidth}
\centering
\includegraphics[scale=0.35]{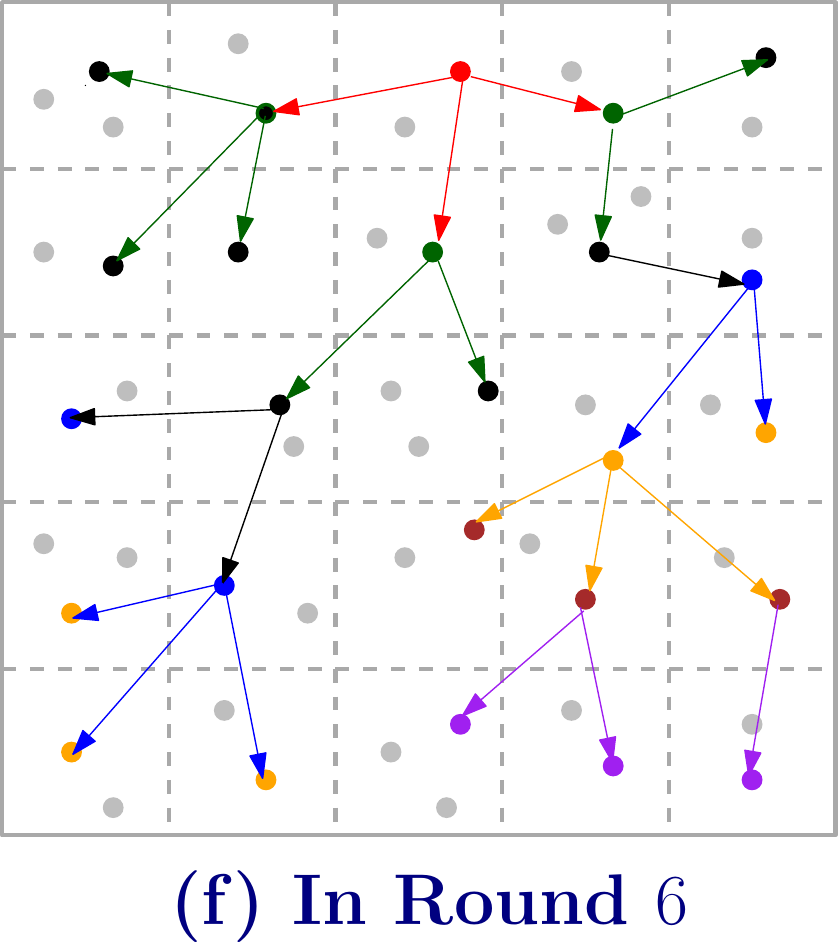}
\end{minipage}
 \caption{steps for making connected set cover}
 \label{dis_fig}
\end{figure}
\vspace{-0.5cm}
In round $1$, each leader $l_i\in {\cal L}$ initiates a partition $C_i=\{l_i\}$. In each round, each node $i\in C_i$ prepares 'Selectlist' consisting maximum number of neighbors, each one from an uncovered block. In case a node gets more than one node from the same block, it selects the neighbor with minimum degree $\cal D$. Node-$i$ sends a 'Selectlist' message to its parent if it is a leaf node in $C_i$.
Else, node $j\in C_i$ selects a list of nodes each belonging to uncovered blocks from its own list and from the received 'Selectlist' messages from its children in same partition. Then it sends the combined 'Selectlist' message to the parent node if it is not a leader node.

The leader node finally from these 'Selectlist' messages selects the nodes to be included and sends the 'Selected' message to them.
If a node receives 'Selected' messages, it selects the parent with minimum $\cal D$ and confirms the request by sending a 'Confirm' message to the corresponding leader. The leader includes the node in $C_i$ and broadcasts the 'Include' message to all nodes in $C_i$. On receiving an 'Include($C_i,j$)' message, all nodes $k\in C_i$ include node-$j$ in its partition and make necessary updates.
In each round, this procedure is repeated until either all blocks are covered by a partition $C_i$ , or no neighbors are left for inclusion.

The formal description of the algorithm is given below.
\vspace{-0.5cm}
\begin{algorithm}[ht]
\scriptsize
\SetLine
  \label{algo:partitioning}
% \linesnumbered
\caption{Distributed algorithm for connected set cover partitioning}
\KwIn{1-hop neighbor list of each node $NL(i)$ with degree ${\cal D}$, {\it Block-Id}, {\it Block~status}, $Status$, $Leader Probability: l_{prob}$}
 \KwOut{Partition $C_i$ from leader $l_i$ }

\For {each node-$i$}
{
\If{node-$i$ is a leader}
{
  $C_i\leftarrow\{i\}, parent=\phi, status=1$\;
}
\If{$Status=1$ and not all blocks covered }
{
	Select neighbors from uncovered blocks and received 'Selectlist' messages, selecting nodes with minimum $\cal D$\;
	\eIf{ $Parent \neq \phi$}
	{
	Send 'Selectlist' to the parent node\;
	}
	{
	  \eIf{$'Selectlist'=\phi$}
	  {
	    Broadcast success=0 and terminate\;
	  }
	  {
	    Send 'Selected' message to the selected neighbors\;
	  }
	}
}
\If{$Status=0$ receives 'Selected' message }
{
	Select that partition where the sender have minimum $\cal D$ and send 'Confirm' message to leader\;
	Update $NL(i)$, {\it Block~status}, $Status$ and include in $C_i$\;
}

\eIf{leader node and receives 'Confirm' message }
{
	Broadcast 'Include' message to all nodes in $C_i$ and update $NL(i)$ , {\it Block~status}, $Status$\;
	\If { all blocks are covered}
	{
	    Broadcast success=1 and terminate\;
	}
}
{
	update $NL(i)$\;
}
 }

 \normalsize
 \end{algorithm}
\vspace{-0.9cm}
\example In Fig.\ref{dis_fig}, it is shown that in round $1$, the leader node (red) selects the neighbors (the green ones) from uncovered blocks. In the next round all black nodes are selected by the red and green nodes. This procedure is repeated to include brown and blue nodes until all blocks are covered. In the last round, all purple nodes are selected and the process is terminated as no uncovered block exists.

It is clear that in each round of the procedure, the nodes already in partition includes several neighbors in the partition so that the partition remains connected with new nodes covering additional blocks. Hence, the procedure terminates faster, each leader either results a successful partition satisfying the condition of connectedness and coverage or it reports a failure when the nodes in the incomplete partition declare them as free nodes.
\vspace{-0.6cm}
\subsection{Distributed Fault Recovery Algorithm}
 \vspace{-0.2cm}
As it has been mentioned in Section 1, once deployed the sensor nodes may fail due to low energy, hardware degradation, inaccurate readings, environmental changes etc.
This paper focuses on the fault recovery problem in case of a single unpredictable node fault in a partition.
It is assumed that when an active node $f$ of a partition $C_i$ fails abruptly, its parent if exists and the children in $C_i$ can detect it.
\vspace{-0.5cm}
\begin{figure}
\begin{minipage}[b]{0.3\linewidth}
\centering
\includegraphics[scale=0.4]{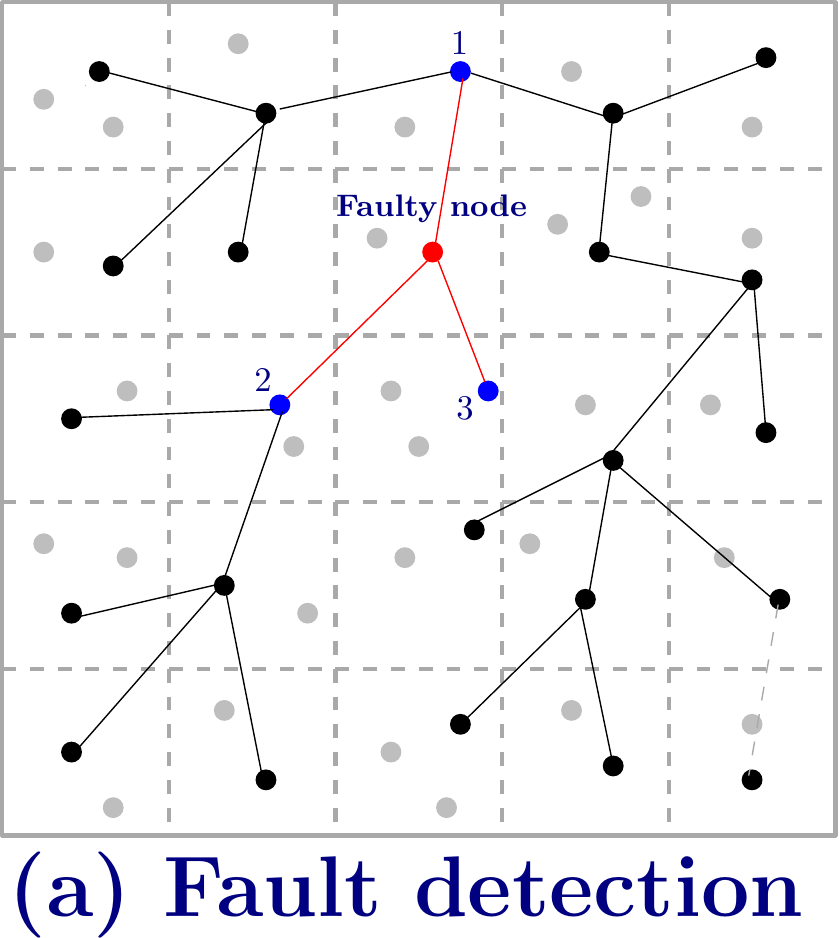}
\end{minipage}
\hspace{0.4cm}
\begin{minipage}[b]{0.3\linewidth}
\centering
\includegraphics[scale=0.4]{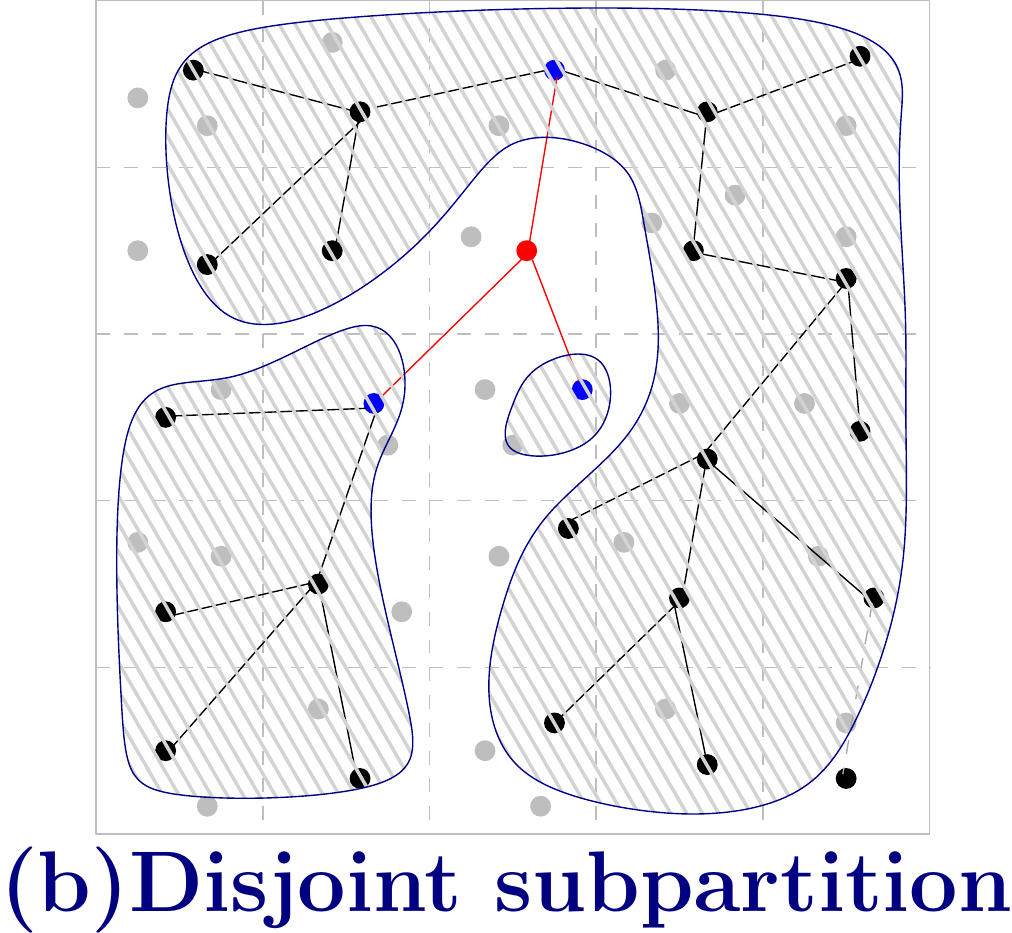}
 \end{minipage}
 \hspace{0.4cm}
 \begin{minipage}[b]{0.3\linewidth}
\centering
\includegraphics[scale=0.4]{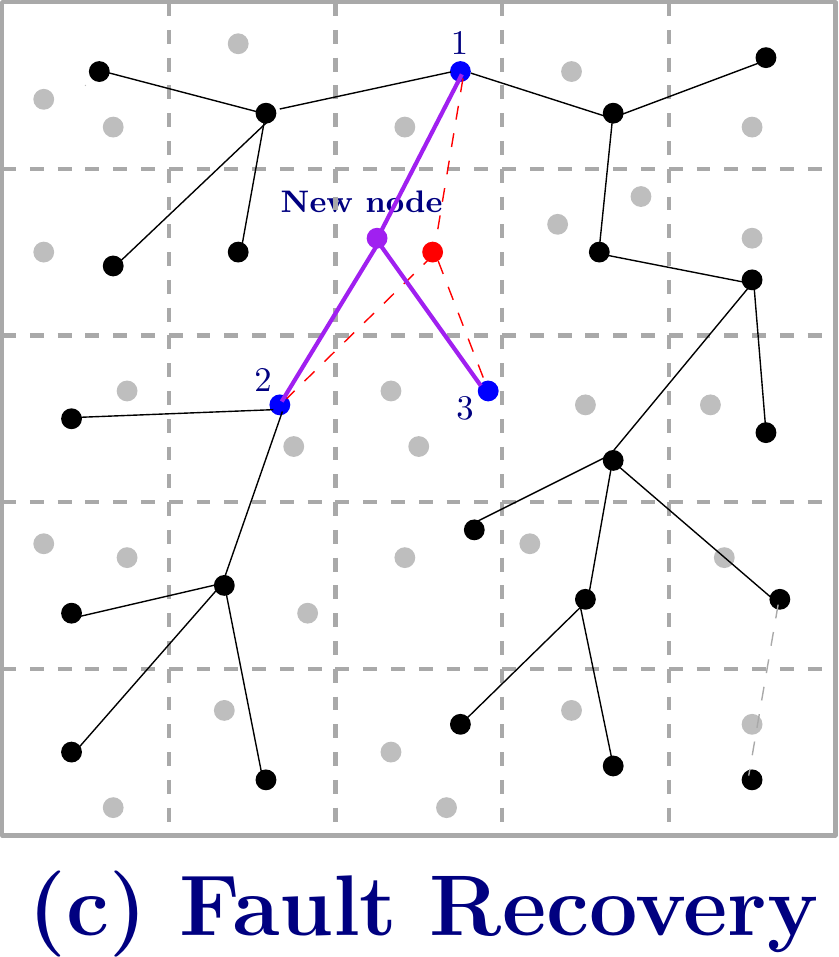}
 \end{minipage}
 \caption{Fault Detection and Recovery Example}
 \label{fig_fault}
\end{figure}
 \vspace{-0.2cm}
A fast fault tolerant algorithm by which all children and the parent of the faulty node in the partition after detecting the faulty node rearrange the partition quickly to make the partition connected the full coverage. If the faulty node is the leader node, its children with the minimum node-id becomes the new leader otherwise the parent node becomes the leader and the fault recovery procedure is initiated by the new leader.\\
The formal description of the algorithm is given below.\\

\begin{algorithm}[H]
\label{fault_algo}
\scriptsize
% \SetLine
%  \linesnumbered
\caption{Distributed fault recovery algorithm}
 \KwIn{faulty node-id : $f$, partition: $C_i$, parent \& children of $f$: $S_i$}
 \KwOut{Recovered partition $C_i$}
 If $f$ is a leader node then $leader_{temp}\leftarrow$ Minimum ID child of $f$, otherwise $leader_{temp}\leftarrow$ parent of $f$\;
 Make a block list $B_S[~]$ covered by $(f\cup S_i)\setminus leader_{temp}$ with $status\leftarrow 0$ \;
 include  $leader_{temp}$ in $temp[~]$\;

 \For { each node-$i\in temp[~]$} {
find neighbors in $S_i$ or from uncovered block. If none, select the free neighbor with maximum $\cal D$\;
Include neighbors in 'Selectlist', \;\
\eIf { node-$i$ not $leader_{temp}$}
{
Send 'Selectlist' message to $leader_{temp}$\;
}
        
      {
            if node-$i$ receives 'Selectelist' message from all nodes-$j\in temp[~]$ ,
             Include nodes in  $temp[~]$\;
              Update $B_S[~]$\;
          
      }

      \If{  $B_S[~] == \phi$}
      {
             Send 'FaultRecovered' message with $temp[~]$ to $\forall j\in C_i$ and terminate\;

      }
      \eIf{ $temp[~]=\phi$}
      {
 	Broadcast 'RecoveryFailed' and terminate\;
      }
      {
    send $temp[~]$ and $B_S[~]$ to all node in $temp[~]$\;

      }

}

 \normalsize
 \label{algo:fault}
 \end{algorithm}
 \vspace{-0.3cm}

\example In Fig. \ref{fig_fault}, say, the red node is faulty. It will be detected by all its children and parent (colored by blue). Now the blue nodes execute the distributed fault recovery algorithm \ref{fault_algo}. After the fault, the partition is broken into three disjoint components as shown in Fig. \ref{fig_fault}(b). The purple node from the faulty node's block is chosen next for maintaining the coverage and it is also connected to at least one node from the disjoint components. Therefore, the connectivity is preserved. Now the new partition including the purple node, is ready for monitoring the area.
% \vspace{-0.2cm}
%  \subsection{ Complexity Analysis}
%  It is evident that the procedure is executed by the children or parent of the faulty node in the same partition. If a node selects a neighbor from the faulty node's block, the coverage criterion is satisfied. All nodes in the partition remain connected except all children or the parent of the faulty node. Therefore, the connectivity can be established by connecting the children or by selecting a new neighbor from the children' blocks. As we have mentioned earlier, within a block all nodes are connected. Therefore, connectivity also is preserved.
% Temporary leader selection in this procedure takes {\emph O($\cal D$)}. Making changes in block status list $B[]$ takes {\emph O($P$)}. Selecting neighbors for the $B[]$ takes {\emph O($P*{\cal D}$)}. After recovering the partition, each node updates their neighbor list in {\emph O($\cal D$)}, Block status list in {\emph O($P$)}, and partition list in {\emph O($P$)}.
% The worst case time complexity is {\emph O($\cal D$)}+{\emph O($P$)}+{\emph O($P*{\cal D}$)}+{\emph O($\cal D$)}+{\emph O($P$)}+{\emph O($P$)} = {\emph O($P*{\cal D}$)}, where $P$ is the number of blocks and $D$ is the maximum degree of the nodes.
% \vspace{-0.5cm}
\vspace{-0.4cm}
\section{Simulation Results and Discussion}
\vspace{-0.3cm}
\label{sec:results}
% \vspace{-0.2cm}
For the simulation studies, we have used network simulator NS 2.34 to evaluate the performance of our proposed algorithm. We have compared our results with \cite{dibakar} that shows significant improvement on the number of rounds for generating the partitions, the network diameter and the number of transmitted messages per node during the procedure.
 \vspace{-0.8cm}
% \vspace{-0.8cm}
\begin{figure}
\begin{minipage}[]{0.45\textwidth}
\centering
\includegraphics[scale=0.30]{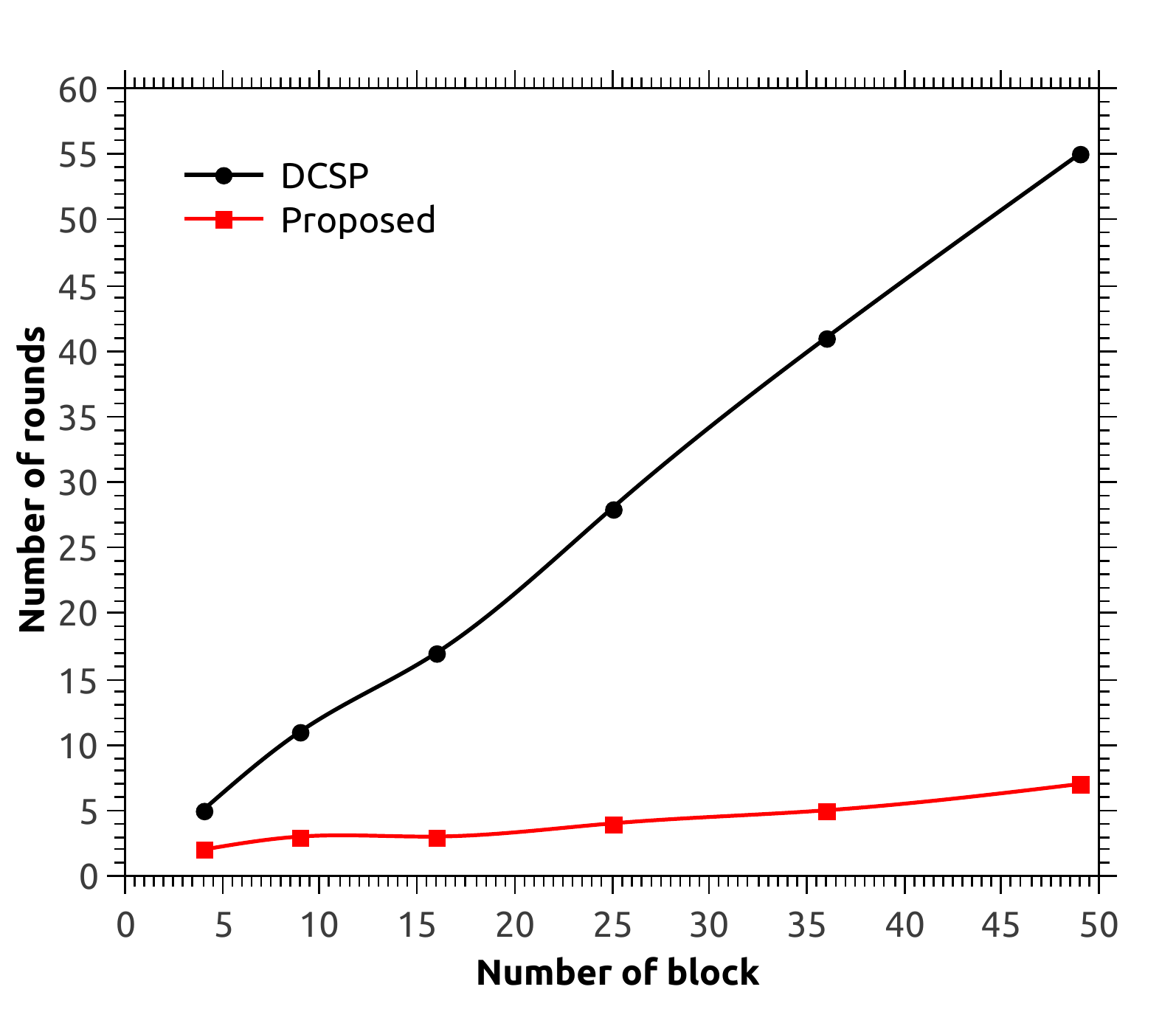}
 \caption{Comparison between DCSP and the Proposed distributed algorithm in terms of average number of rounds for partitioning}
 \label{result1}
\end{minipage}
\hspace{0.6cm}
\begin{minipage}[]{0.47\textwidth}
\centering
\includegraphics[scale= 0.30]{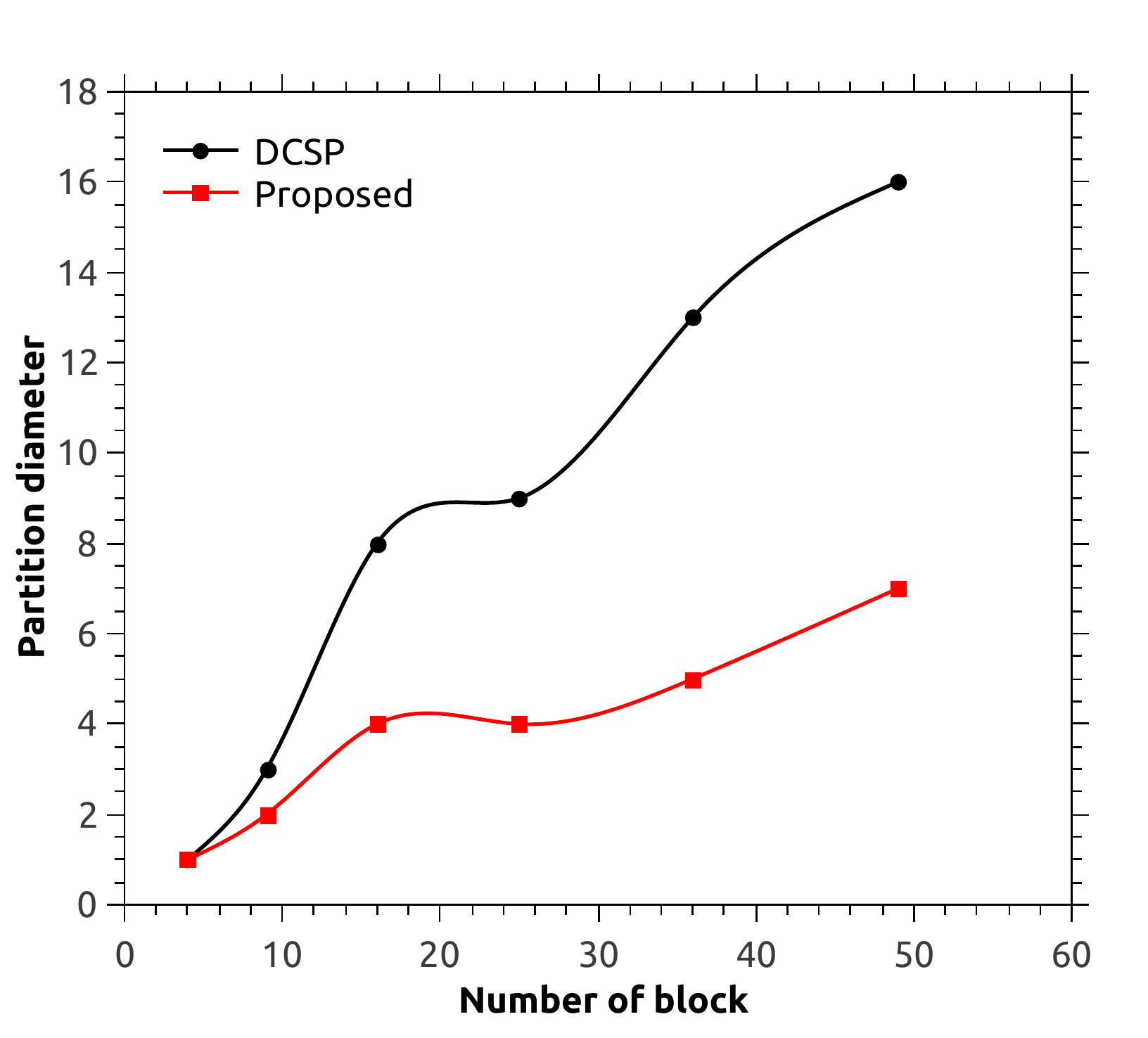}
 \hspace{0.8cm}
 \caption{Comparison between DCSP and the Proposed distributed algorithm in terms of partition diameter}
 \label{result2}
 \end{minipage}
\end{figure}
\vspace{-0.6cm}
The sensor nodes are deployed over a grid $P$ which is divided into a number of blocks ($2 \times 2$),( $3 \times 3$) to ( $7 \times 7$) respectively. Fig.\ref{result1} shows the variation of the average number of rounds to complete partitioning with the grid size. Obviously, the number of rounds increases with the number of blocks. However, compared to the {\it DCSP algorithm} proposed in \cite{dibakar}, the present method completes in significantly less number of rounds. Therefore, during initialization, the proposed method will converge faster to achieve the connected covers of the nodes.
\vspace{-0.1cm}
\begin{figure}[!ht]
\begin{minipage}{0.435\textwidth}
\includegraphics[scale=0.25]{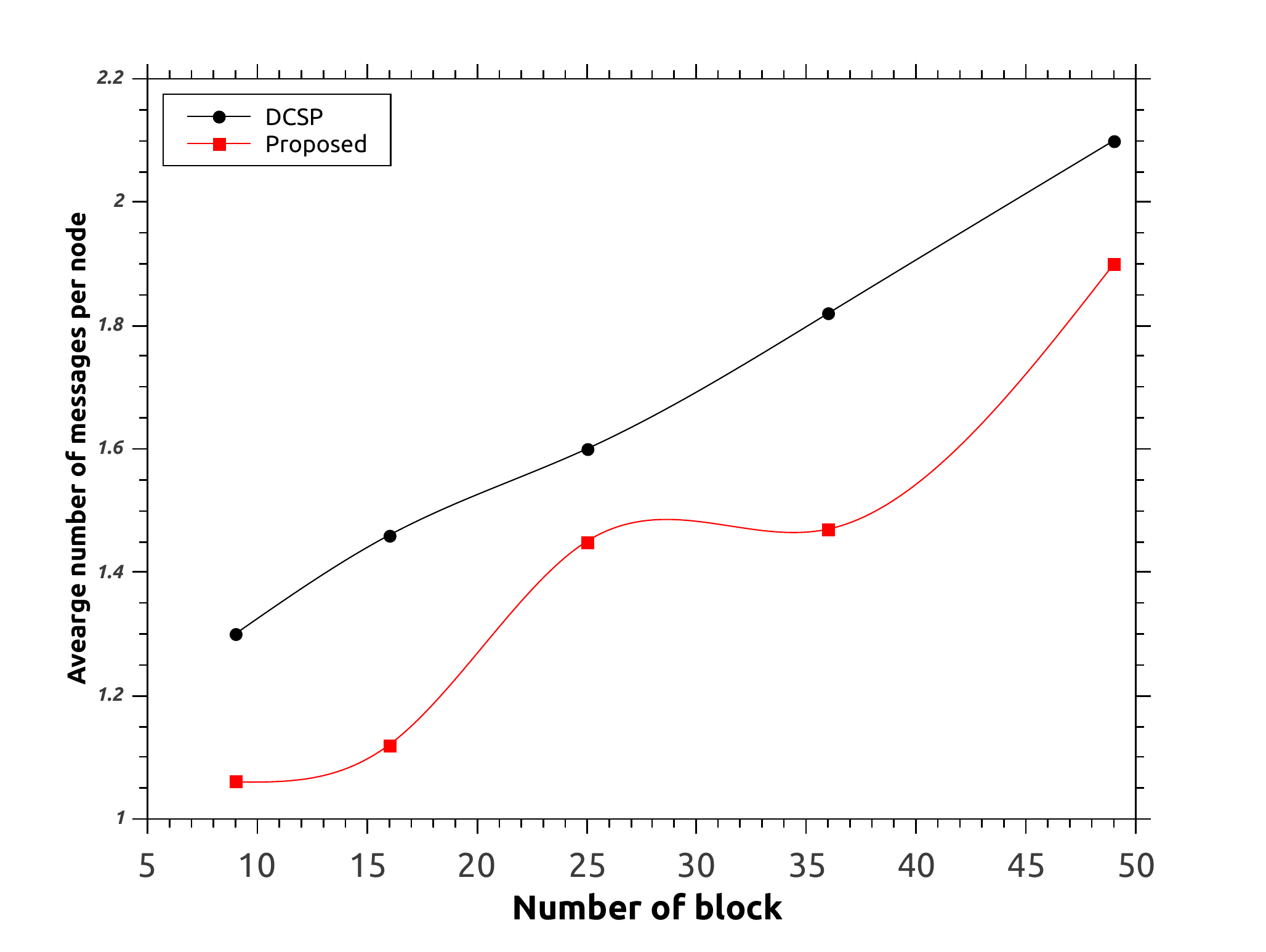}
 \caption{DCSP vs Proposed distributed algorithm in terms of average number of transmitted messages }
 \label{result3}
 \end{minipage}
  \hspace{0.5cm}
 \begin{minipage}{0.48\textwidth}
\includegraphics[keepaspectratio=true,scale=0.205]{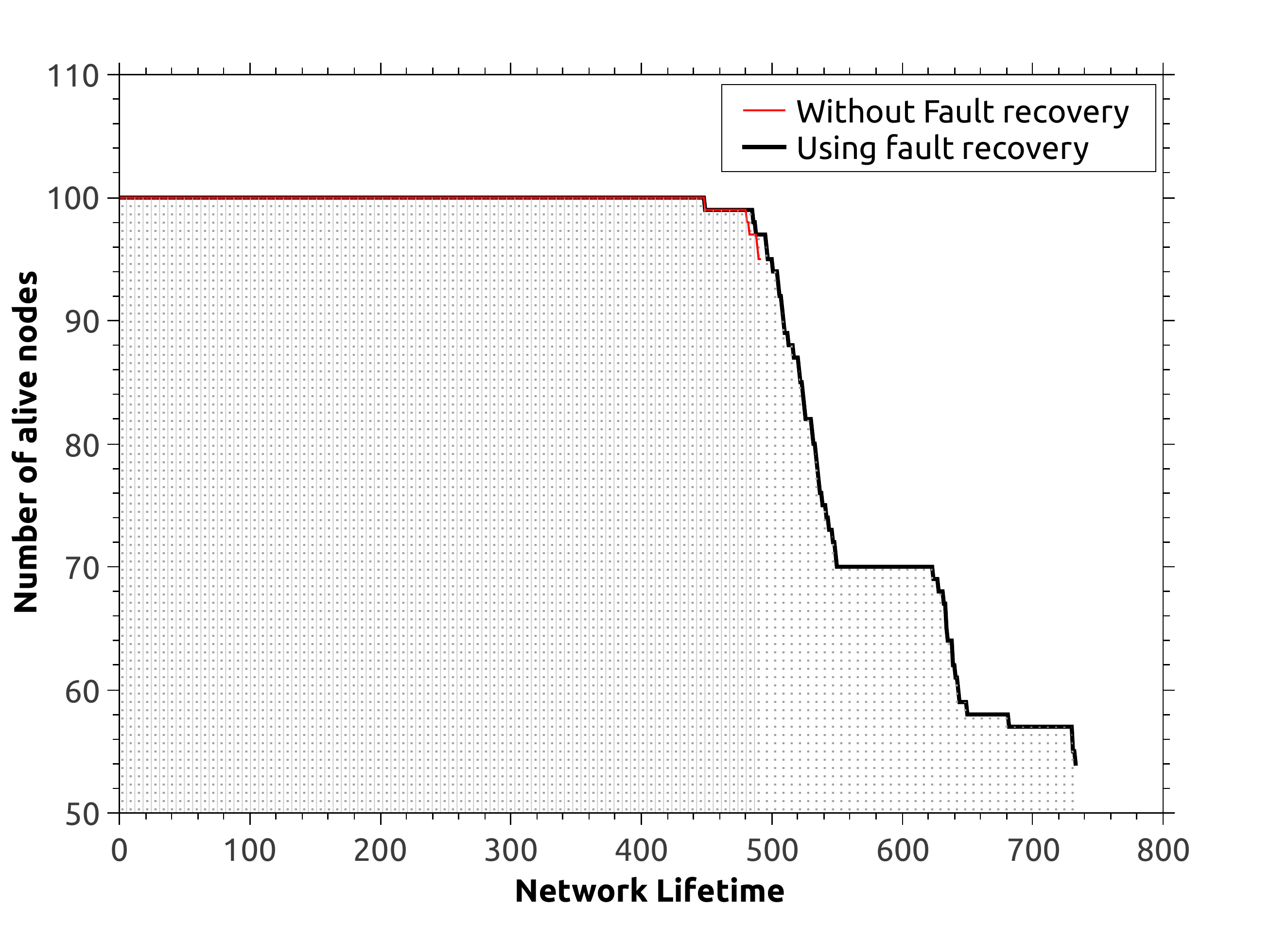}
 \caption{Extension of network lifetime using fault recovery technique vs without fault recovery }
 \label{lifetime}
 \end{minipage}
 \end{figure}

In Fig.\ref{result2}, it is shown that the proposed algorithm results significant improvement in terms of network diameter over the DCSP algorithm\cite{dibakar}. In a network with large diameter, the number of steps to route a message from a source node to a destination node will require more delay and more communications between intermediate nodes.
Therefore, the low diameter network topologies are preferred for a partition that can aggregate the data using less number of hops, i.e., with less delay and less number of broadcasts.

Also, Fig. \ref{result3} shows the significant improvement in average number of transmitted messages per node in computing the connected set covers. Since the procedure terminates faster using fewer rounds of computation, the total number of messages exchanged per node is also less here.
Finally, Fig. \ref{lifetime} shows how the fault recovery algorithm enhances the network lifetime in presence of faults. Though the proposed fault model includes any unpredicted node faults, in the simulation, only node faults due to energy exhaustion has been taken into account. Simulation results show almost $50\%$ enhancement in network lifetime.
% \vspace{-0.3cm}
\vspace{-0.4cm}
\section{Conclusion}
\vspace{-0.3cm}
% \vspace{-0.3cm}
\label{sec:conclusion}
In this paper, we have focused on the connected set cover partitioning problem. A self-organized fast distributed algorithm is proposed for finding maximum number of connected cover partitions. Also, distributed fault recovery technique is developed to rearrange connected set covers in presence of unpredictable node faults to satisfy both connectivity and coverage criteria. Minimization of network diameter of the partition and significant improvements in terms of computation rounds and message overhead are also achieved by our proposed method.
In summary, the proposed connected set cover partitioning technique along with the localized fault recovery scheme opens up new avenues for setting up self organized wireless sensor networks with enhanced lifetime.

\bibliographystyle{splncs}
% \balancecolumns
\vspace{-0.52cm}
\bibliography{refferences}

\end{document}